\documentclass[aps,prl,twocolumn,superscriptaddress]{revtex4-1}

\usepackage{graphicx,latexsym}
\usepackage{amsmath,amssymb,amsfonts}
\usepackage{bm}
\usepackage{braket}

\begin{document}


\title{Non-Hermitian Kondo Effect in Ultracold Alkaline-Earth Atoms}
\author{Masaya Nakagawa}
\email{masaya.nakagawa@riken.jp}
\affiliation{RIKEN Center for Emergent Matter Science (CEMS), Wako, Saitama 351-0198, Japan}
\author{Norio Kawakami}
\affiliation{Department of Physics, Kyoto University, Kyoto 606-8502, Japan}
\author{Masahito Ueda}
\affiliation{Department of Physics, University of Tokyo, 7-3-1 Hongo, Bunkyo-ku, Tokyo 113-0033, Japan}
\affiliation{RIKEN Center for Emergent Matter Science (CEMS), Wako, Saitama 351-0198, Japan}




\date{\today}

\begin{abstract}
We investigate the Kondo effect in an open quantum system, motivated by recent experiments with ultracold alkaline-earth(-like) atoms. 
Because of inelastic collisions and the associated atom losses, this system is described by a complex-valued Kondo interaction and provides a non-Hermitian extension of the Kondo problem. 
We show that the non-Hermiticity induces anomalous \textit{reversion} of renormalization-group flows which violate the $g$-theorem due to non-unitarity and produce a quantum phase transition unique to non-Hermiticity. 
Furthermore, we exactly solve the non-Hermitian Kondo Hamiltonian using a generalized Bethe ansatz method and find the critical line consistent with the renormalization-group flow.
\end{abstract}


\maketitle



Isolated quantum systems are governed by unitary dynamics and described by Hermitian Hamiltonians, yet
no quantum system is completely isolated in reality and dissipation is ubiquitous in nature. 
The non-unitary dynamics of open quantum systems permits an effective description based on \textit{non-Hermitian} Hamiltonians under an appropriate condition \cite{Daley14, Dalibard92}.  
Contrary to the conventional wisdom that the dissipation is detrimental to quantum coherence, studies of non-Hermitian quantum dynamics have revealed unique quantum phenomena such as unconventional phase transitions from real to complex energy spectra \cite{Bender98, Bender07, Heiss12}, quantum critical behavior beyond the equilibrium universality class \cite{Ashida16, Ashida17, Kawabata17}, and exotic topological phases \cite{Rudner09, Lee16,Shen18, Kunst18, Yao18, Kawabata18_3, Gong18, Kawabata18_2}. Experiments on these phenomena have rapidly progressed over the past decade using engineered dissipation in optical systems and ultracold atoms \cite{Guo09, Ruter10, Feng11, Regensburger12, Zeuner15, Li16, Xiao17, Zhou18}. 

However, most of the previous studies focused on single-particle quantum mechanics, and many-body physics with interparticle interactions has not been explored barring some exceptions \cite{Durr09, GarciaRipoll09, Daley09, Kantian09, Ashida16, Ashida17}. 
In fact, many-body systems exhibit emergent behavior which cannot be explained by a simple single-particle picture.  
If the interactions are arbitrarily weak, their effects can be significant and even non-perturbative, as represented by the BCS theory of superconductivity \cite{Coleman_book}. 
Therefore, the interplay between strong correlations and non-Hermiticity is expected to bring about hitherto unnoticed quantum many-body effects inherent in open quantum systems.

\begin{figure}
\includegraphics[width=8.5cm]{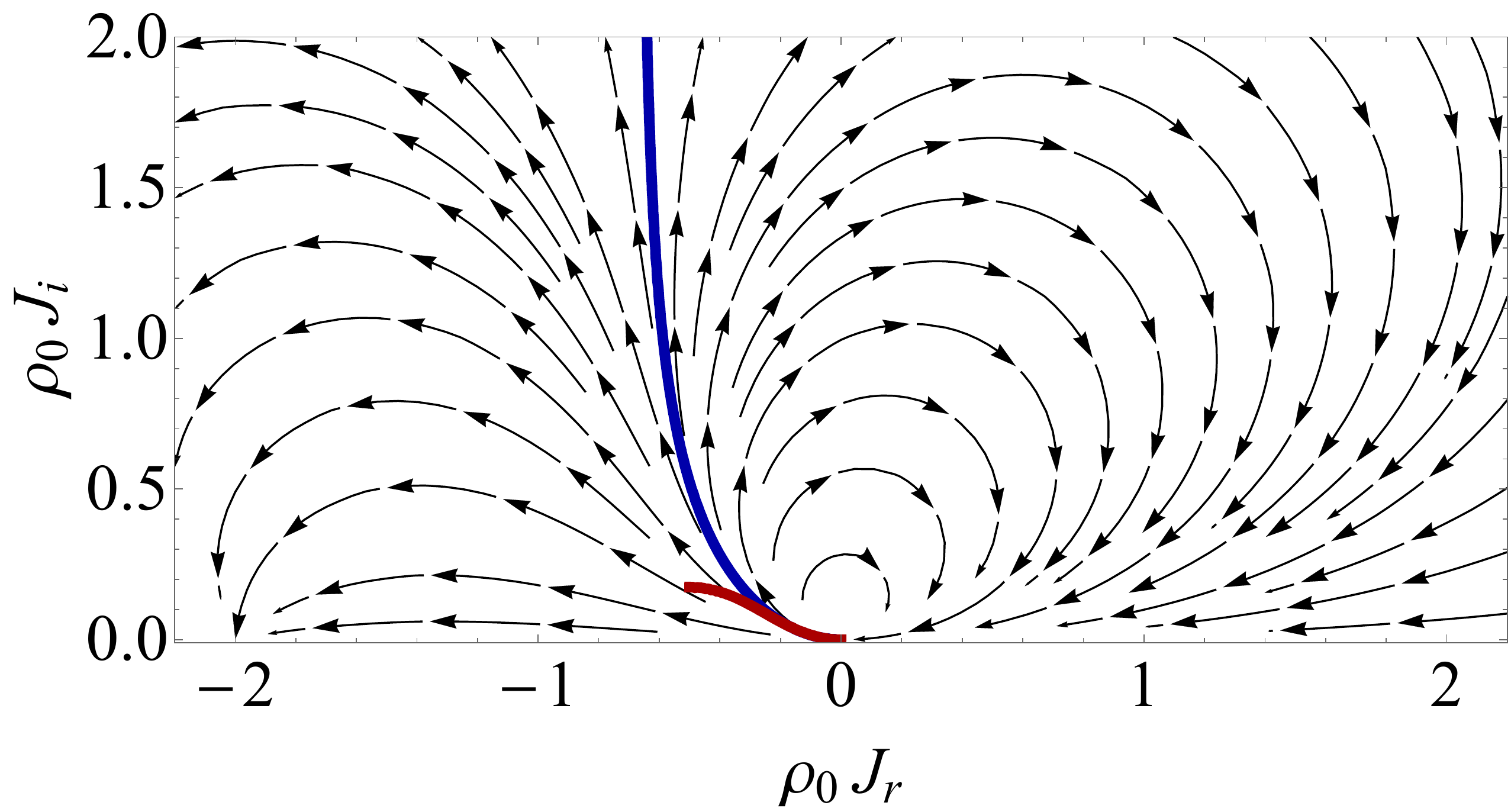}
\caption{RG flow of the non-Hermitian Kondo model \eqref{eq_Heff} up to the 2-loop order. The blue curve shows the critical line obtained from the analytical solution of the RG equation (Eq.\ \eqref{eq_RGcrit} in Supplemental Material \cite{supple}), and the red curve is the critical line obtained from the Bethe-ansatz solution (Eq.\ \eqref{eq_transition}).}
\label{fig_RGflow}
\end{figure}

In this Letter, we study a quantum many-body effect in a non-Hermitian interacting  system, highlighting the role of interactions with \textit{complex} coefficients. 
Our focus is a paradigmatic Fermi-surface effect in strongly correlated systems: the Kondo effect \cite{Coleman_book, Kondo64, Hewson}. 
This effect serves as a minimal physical setup to investigate the strong correlation caused by a single magnetic impurity immersed in a Fermi sea. 
At low temperatures, low-energy excitations near the Fermi surface cooperatively form a many-body spin-singlet state with the impurity, and this Kondo singlet exhibits a non-perturbative energy dependence on the interaction. 
We show that a recent experimental realization of the Kondo system with ultracold atoms \cite{Riegger18} offers a \textit{non-Hermitian Kondo Hamiltonian} due to inelastic collisions and the associated atom losses, thereby generalizing the Kondo problem to non-Hermitian physics. 
Employing the Kondo Hamiltonian with complex-valued interactions, we find that the non-Hermiticity induces an exotic renormalization-group (RG) flow where \textit{the flow starting from a fixed point eventually returns to the original point} (see Fig.\ \ref{fig_RGflow}). 
Such reversion of RG flows manifestly violates the $g$-theorem \cite{AffleckLudwig91, Friedan04}, presenting a spectacular physical consequence of non-Hermiticity. 
We also find a quantum phase transition between the Kondo phase and the non-Kondo phase, accompanied by divergence of the non-Hermitian interaction at the critical point.

Moreover, we find an exact solution of this non-Hermitian Kondo problem by using a generalized Bethe ansatz method \cite{Andrei80, Wiegmann81, Andrei83, TsvelickWiegmann83}, which demonstrates that the integrability of the Kondo model is not spoiled even if the interaction coupling constant is complex. Thus our model affords a nontrivial many-body example of non-Hermitian quantum integrable models. 
The obtained exact result for the critical line shows a good agreement with the prediction of the RG. 

\textit{Setup}.--\ 
We first describe our setup and derive the non-Hermitian Kondo Hamiltonian. Our setup is similar to the recent experiment using ultracold alkaline-earth-like atoms \cite{Riegger18}. We consider an equilibrium gas of alkaline-earth-like fermionic atoms in the electronic ground state ($^1S_0$) in a three-dimensional optical lattice. We assume that the atoms partially occupy the lowest band of the tight-binding model and thus form a metallic state. Then, a weak laser, which is tuned for the clock transition, excites a fraction of atoms to a metastable excited state ($^3P_0$). By choosing an appropriate optical lattice wavelength, the atoms in the $^3P_0$ state can strongly be confined and behave as immobile impurities, whereas those in the $^1S_0$ state can move between lattice sites \cite{Gorshkov}. Since both of the electronic states have nuclear spin degrees of freedom (here we assume spin $1/2$), the system around an impurity is described by the Kondo Hamiltonian \cite{Gorshkov}
\begin{align}
H&=\sum_{\bm{k},\sigma}\varepsilon_{\bm{k}}c_{\bm{k}\sigma}^\dag c_{\bm{k}\sigma}\notag\\
&+\frac{1}{N_s}\sum_{\bm{k},\bm{k}',\sigma,\sigma'}c_{\bm{k}\sigma}^\dag c_{\bm{k}'\sigma'}(v_r\delta_{\sigma\sigma'}-J_r\bm{\sigma}_{\sigma\sigma'}\cdot\bm{S}_{\mathrm{imp}}).
\label{eq_Kondo}
\end{align}
Here, $c_{\bm{k}\sigma}$ denotes the annihilation operator of the $^1S_0$ atoms with momentum $\bm{k}$ and spin $\sigma=\uparrow,\downarrow$, $\varepsilon_{\bm{k}}$ is the band dispersion, and $N_s$ is the number of sites. The last two terms in Eq.\ \eqref{eq_Kondo} describe 
the interactions between free fermions and the impurity, where $\bm{\sigma}$ is the three-component Pauli matrix vector and $\bm{S}_{\mathrm{imp}}$ is the impurity spin operator. The spin-independent potential scattering $v_r$ and the spin-exchange scattering $J_r$ are related to the $s$-wave scattering lengths $a_{eg}^+$ ($a_{eg}^-$) in the spin-singlet (triplet) channel as $v_r\propto a_{eg}^++3a_{eg}^-$ and $J_r\propto a_{eg}^+-a_{eg}^-$ \cite{Gorshkov} (see also Ref.\ \cite{KanaszNagy18}). 

The Kondo effect in ultracold alkaline-earth-like atoms has been extensively studied in literature \cite{Gorshkov, FossFeig1, FossFeig2, Silva-Valencia12, Isaev15, Nakagawa15, Zhang16, Kuzmenko16, Cheng17, Nakagawa17, Kuzmenko18, KanaszNagy18}. However, the previous studies did not consider the inelastic scattering between the $^1S_0$ and $^3P_0$ states, which causes two-body losses of scattered atoms as observed experimentally \cite{Scazza, Hofer, Pagano15, Riegger18}.  
As time elapses, some of the impurities in the initial state are lost due to inelastic collisions but other impurities will survive. The atom losses are described by a quantum master equation \cite{Daley14} 
\begin{align}
\frac{d\rho(t)}{dt}&=-i[H,\rho]+\sum_{\alpha=+,-,\uparrow\uparrow,\downarrow\downarrow}(L_\alpha\rho L_\alpha^\dag-\frac{1}{2}\{ L_\alpha^\dag L_\alpha,\rho\})\notag\\
&=-i(H_{\mathrm{eff}}\rho-\rho H_{\mathrm{eff}}^\dag)+\sum_\alpha L_\alpha\rho L_\alpha^\dag,
\label{eq_master}
\end{align}
where $\rho(t)$ is the density matrix of the atomic cloud. The Lindblad operators $L_\pm, L_{\uparrow\uparrow}, L_{\downarrow\downarrow}$ describe the two-body losses of $^1S_0$ and $^3P_0$ atoms via the corresponding inelastic scattering channels in spin states $\ket{\pm}=(\ket{\uparrow\downarrow}\pm\ket{\downarrow\uparrow})/\sqrt{2}, \ket{\uparrow\uparrow}, \ket{\downarrow\downarrow}$ (see Ref.\ \cite{supple} for their explicit forms). 
Such two-body losses emerge as effective imaginary interactions in the non-Hermitian Hamiltonian $H_{\mathrm{eff}}=H-\frac{i}{2}\sum_{\alpha=+,-,\uparrow\uparrow,\downarrow\downarrow}L_\alpha^\dag L_\alpha$. 
By unraveling the dynamics of the density matrix into quantum trajectories \cite{Daley14, Dalibard92}, we can decompose the dynamics into the Schr\"{o}dinger evolution under the effective non-Hermitian Hamiltonian 
and a stochastic quantum-jump process described by the last term in the second line of Eq.\ \eqref{eq_master}. Note that the quantum jumps cause the loss of impurity atoms from the trap; therefore, the dynamics around a surviving impurity is obtained by projecting out the quantum jumps and described by the non-Hermitian Kondo Hamiltonian
\begin{align}
H_{\mathrm{eff}}&=\sum_{\bm{k},\sigma}\varepsilon_{\bm{k}}c_{\bm{k}\sigma}^\dag c_{\bm{k}\sigma}\notag\\
&+\frac{1}{N_s}\sum_{\bm{k},\bm{k}',\sigma,\sigma'}c_{\bm{k}\sigma}^\dag c_{\bm{k}'\sigma'}(v\delta_{\sigma\sigma'}-J\bm{\sigma}_{\sigma\sigma'}\cdot\bm{S}_{\mathrm{imp}})
\label{eq_Heff}
\end{align}
with complex-valued interactions $v=v_r+iv_i$ and $J=J_r+iJ_i$ ($v_r,v_i,J_r,J_i\in\mathbb{R}$) \cite{supple}. After the excitation of the $^3P_0$ state, the atomic gas around the impurity undergoes the quench dynamics under $H_{\mathrm{eff}}$. We note that, even if there is no loss event at the impurity site, the effect of inelastic scattering is not negligible; the backaction from projecting out quantum jumps influences the behavior of the system through the non-Hermitian part of $H_{\mathrm{eff}}$. In this Letter, we analyze the properties of $H_{\mathrm{eff}}$ and focus on whether or not the eigenstates show the Kondo effect.


\textit{Renormalization-group analysis}.--\ 
To unveil the Kondo physics in the non-Hermitian Hamiltonian \eqref{eq_Heff}, we first employ the poor-man's RG method \cite{Anderson70} of integrating out the high-energy part of the conduction band. Note that even if the Hamiltonian \eqref{eq_Heff} is non-Hermitian, the dispersion relation $\varepsilon_{\bm{k}}$ of the conduction band is real and thus the high-energy part is well defined. Since the poor-man's scaling can formally be performed regardless of whether the coupling $J$ is real or complex, we obtain the RG equation up to the 2-loop order which takes the same form as in the Hermitian case \cite{NozieresBlandin, footnote_perturbativeRG}:
\begin{equation}
\frac{dJ}{d\ln D}=\rho_0J^2+\frac{\rho_0^2}{2}J^3,
\label{eq_RG}
\end{equation}
where $D$ is one-half of the bandwidth of the conduction band and $\rho_0$ is the density of states at the Fermi energy. For simplicity, here we have neglected the potential scattering since it does not affect the qualitative behavior as shown later. 
Figure \ref{fig_RGflow} shows the RG flow in the complex-interaction plane. On the real axis (the Hermitian Kondo problem), the system flows from the free fixed point $J=0$ to the Kondo fixed point $J=-2/\rho_0$. Remarkably, the RG flow extended to the non-Hermitian case indicates a quantum phase transition between the Kondo phase and the non-Kondo phase separated by a critical line (blue curve in Fig.\ \ref{fig_RGflow}). 
An analytical formula for the critical line is available in the Supplemental Material \cite{supple}. On the critical line, the imaginary part of the coupling diverges at $J_r=-2/(3\rho_0)$. 

We emphasize that the phase transition from the Kondo phase to the non-Kondo phase should not be regarded as a consequence of decoherence due to the atom loss, since no atom is lost at the surviving impurity site. 
The physical origin of the transition is attributed to a phenomenon similar to the continuous quantum Zeno effect \cite{Syassen08, Mark12, Barontini13, Zhu14, Tomita17}; 
the strong losses effectively deplete particles surrounding the impurity, thereby destroying the Kondo singlet. 
Since the Kondo singlet is formed in the spin sector, the phase transition cannot be caused by the inelastic potential scattering (which only affects the charge sector); it requires the imaginary spin-exchange interaction.

Furthermore, the RG flow shown in Fig.\ \ref{fig_RGflow} has a dramatic feature. In the non-Kondo phase with $J_r<0$, the RG flow starts from the free fixed point and eventually returns back to the original fixed point. Such reversion of the RG flow is usually forbidden in Hermitian cases, since the $g$-theorem \cite{AffleckLudwig91, Friedan04} dictates that the ground-state degeneracy monotonically decrease along the RG flow. In our case, the non-Hermiticity breaks the unitarity, thereby invalidating one of the key assumptions of the $g$-theorem. Thus, the RG flow in Fig.\ \ref{fig_RGflow} is allowed by the non-Hermiticity.

To understand the physics of the reversion of the flows, we calculate an energy scale $T_{\mathrm{Kdiss}}$ defined by $J_r(T_{\mathrm{Kdiss}})=0$ and $J_i(T_{\mathrm{Kdiss}})\neq 0$. This energy scale corresponds to a characteristic scale where the dissipative Kondo system begins to show the reversion of the running coupling constants to the free fixed point. 
As detailed in the Supplemental Material \cite{supple}, the result is
\begin{align}
T_{\mathrm{Kdiss}}=&\frac{D}{\sqrt{2}}\Bigl(1+\frac{4}{(\rho_0\tilde{J}_i)^2}\Bigr)^{\frac{1}{4}}\Bigl|\frac{\rho_0J}{1+\frac{1}{2}\rho_0J}\Bigr|^{\frac{1}{2}}
\exp\left[\frac{J_r}{\rho_0|J|^2}\right],
\label{eq_TKdiss}
\end{align}
where $\tilde{J}_i$ is the imaginary Kondo coupling at that scale. 
Near the critical line and for $|\rho_0J|\ll 1$, this expression is simplified as
\begin{align}
T_{\mathrm{Kdiss}}\simeq \frac{D}{\sqrt{2}}\sqrt{|\rho_0J|}
\exp\left[\frac{J_r}{\rho_0|J|^2}\right],
\end{align}
which is a natural generalization of the well-known form of the Kondo temperature \cite{Coleman_book} to the non-Hermitian case. Thus, the reversion of the RG flows is non-perturbative in terms of the Kondo coupling. 
The new non-perturbative scale $T_{\mathrm{Kdiss}}$ can be regarded as a remnant of the Kondo physics after the transition into the non-Kondo phase induced by non-Hermiticity.

\begin{figure}
\includegraphics[width=8.5cm]{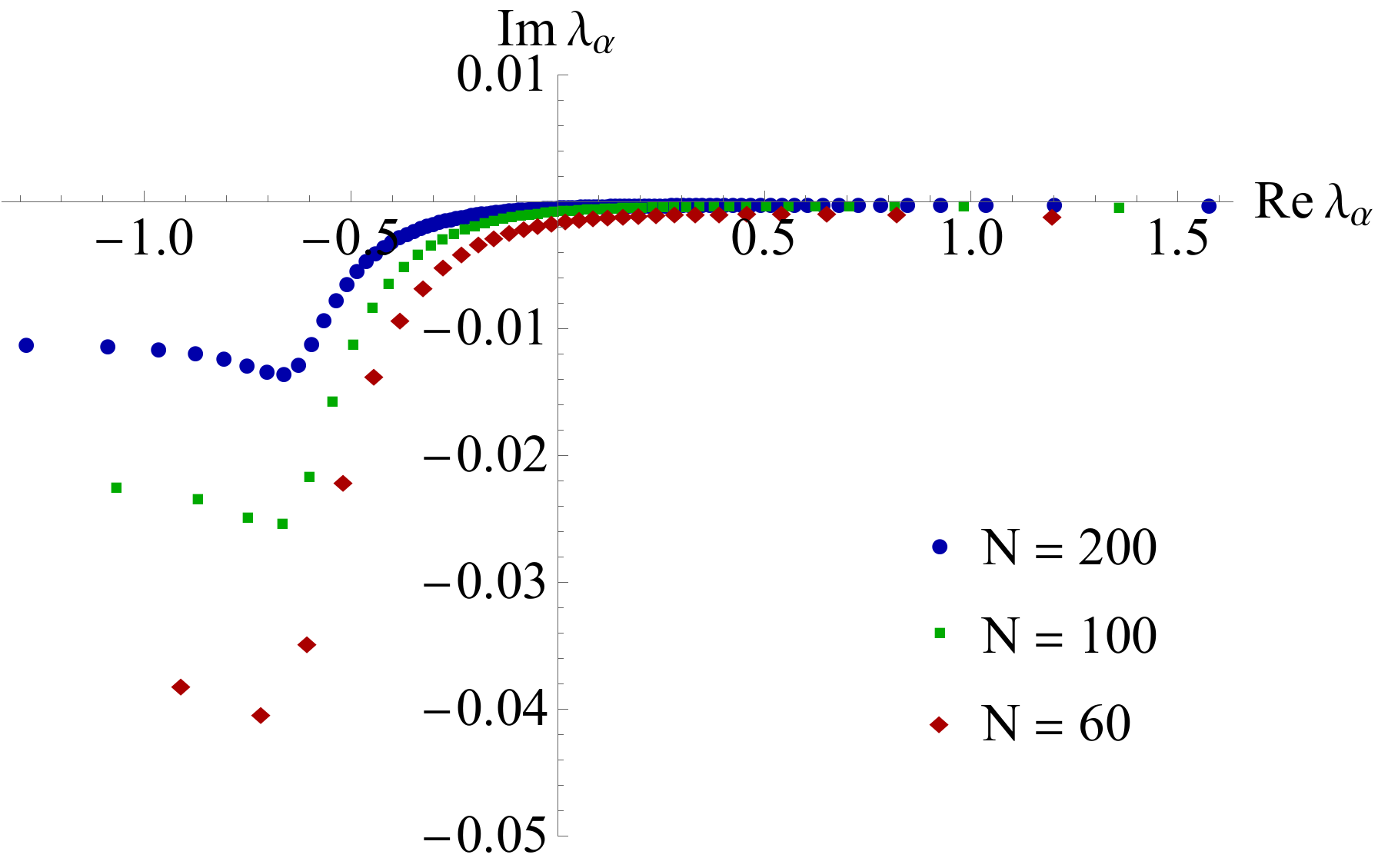}
\caption{Spin rapidities obtained from the numerical solutions of the Bethe equations \eqref{eq_Bethelog2} for the total number of particles $N=60, 100, 200$ and the number of spin-down particles $M=N/2$. The Kondo coupling is set to be $\rho_0J=-0.3+0.1i$.}
\label{fig_Betheroots}
\end{figure}

\textit{Generalized Bethe-ansatz solution}.--\ 
So far the non-Hermitian Kondo physics has been discussed on the basis of the perturbative RG, which is applicable only in the weak coupling regime. To confirm the prediction of the RG flow, we derive an exact solution of the non-Hermitian Kondo model \eqref{eq_Heff} by using the Bethe ansatz method \cite{Andrei80, Wiegmann81, Andrei83, TsvelickWiegmann83}. The low-energy behavior of the Kondo model is exactly solvable if the band dispersion is linearized around the Fermi energy. In the non-Hermitian physics, this low-energy condition is understood as the condition for the real part of the energy. The Yang-Baxter integrability condition for the Kondo model reads
\begin{equation}
P_{12}R_{10}R_{20}=R_{20}R_{10}P_{12},
\end{equation}
where $P_{12}=\frac{1}{2}(1+\bm{\sigma}_1\cdot\bm{\sigma}_2)$ and $R_{j0}=\exp[-2\pi i\rho_0v-i\pi\rho_0 J\bm{\sigma}_j\cdot\bm{S}_{\mathrm{imp}}]$. 
Notably, this Yang-Baxter relation holds for arbitrary $v,J\in\mathbb{C}$; therefore, the integrability of the Kondo model is maintained even if the Kondo interaction is complex. This striking property enables us to obtain exact results for the non-Hermitian Kondo model. The Bethe equations are given by
\begin{gather}
k_jL=2\pi I_j-2\pi\rho_0 v-\pi\rho_0J/2-\sum_{\alpha=1}^M(\theta(\lambda_\alpha)+\pi),\label{eq_Bethelog1}\\
N\theta(\lambda_\alpha)=2\pi K_\alpha-\theta(\lambda_\alpha+1/g)+\sum_{\beta=1}^M\theta(\frac{\lambda_\alpha-\lambda_\beta}{2}),\label{eq_Bethelog2}
\end{gather}
where 
$\theta(x)=2\arctan(2x)$, $g=-\tan(\pi\rho_0J)$, 
$j=1,\cdots,N$, and $\alpha=1,\cdots,M$. 
Here, $k_j$ and $\lambda_\alpha$ denote the quasimomentum and the spin rapidity, respectively, $N$ is the number of the conduction fermions, $M$ is the number of spin-down particles, 
and $L$ is the length of the effective one-dimensional system after the linearization of the dispersion. 
The quantum numbers are taken as $I_j\in\mathbb{Z}\ (\mathbb{Z}+1/2)$ for even (odd) $N$, and $K_\alpha\in\mathbb{Z}\ (\mathbb{Z}+1/2)$ for even (odd) $N-M$. 
Since the effect of the potential scattering $v$ is an overall shift of the quasimomenta, it does not contribute to the Kondo physics which is determined by the spin part \eqref{eq_Bethelog2}. A numerical solution of the Bethe equations \eqref{eq_Bethelog2} is plotted in Fig.\ \ref{fig_Betheroots}. Here we show the solution that is continuously connected to that of the ground state in the Hermitian case by setting $K_\alpha=\frac{N-M}{2}-(\alpha-1)$. Reflecting the non-Hermiticity, the spin rapidity takes complex values in general. However, the deviation from the real axis is small and negligible in the thermodynamic limit $N\to\infty$, since the non-Hermiticity appears only through the impurity part. 
Since the effect of the single impurity becomes irrelevant in the $N\to\infty$ limit in Eqs.\ \eqref{eq_Bethelog1} and \eqref{eq_Bethelog2}, the Kondo physics appears through the $1/N$ correction in the physical quantities calculated from the Bethe-ansatz solution.

Now let us examine the property of the ground state (in the sense of the real part of the energy) from the Bethe equations for the case of $M=N/2$. We introduce the density of the spin rapidities by $\sigma(\lambda)\equiv\frac{1}{N}\frac{d K(\lambda)}{d\lambda}=a_1(\lambda)+\frac{1}{N}a_{1}(\lambda+1/g)-\frac{1}{N}\sum_{\beta=1}^Ma_2(\lambda-\lambda_\beta)$ with $a_n(\lambda)=\frac{1}{2\pi}\frac{d\theta(\lambda/n)}{d\lambda}=\frac{1}{2\pi}\frac{n}{\lambda^2+n^2/4}$. In the thermodynamic limit, we can replace the sum with the integral as $\frac{1}{N}\sum_{\beta=1}^M \to \int_\mathcal{C} d\lambda' \sigma(\lambda')$ and thus obtain an integral equation for $\sigma(\lambda)$:
\begin{align}
\sigma(\lambda)=a_1(\lambda)+\frac{1}{N}a_{1}(\lambda+1/g)-\int_\mathcal{C}d\lambda' a_2(\lambda-\lambda')\sigma(\lambda').
\label{eq_Betheinteg1}
\end{align}
The trajectory $\mathcal{C}$ runs over $(-\infty,\infty)$ in the Hermitian case. 
In the non-Hermitian case, it shows a small detour (of the order of $1/N$) from the real axis, but  can be deformed onto it due to the analyticity of $a_2(\lambda-\lambda')\sigma(\lambda')$. 
To extract the contribution from the impurity, we divide the density into the host part and the impurity part as 
$\sigma(\lambda)=\sigma_h(\lambda)+\frac{1}{N}\sigma_i(\lambda)$. 
Substituting this into Eq.\ \eqref{eq_Betheinteg1} and extracting the $1/N$ term, we obtain
\begin{align}
\sigma_i(\lambda)&=a_{1}(\lambda+1/g)-\int_{-\infty}^\infty d\lambda' a_2(\lambda-\lambda')\sigma_i(\lambda').\label{eq_Betheintegi}
\end{align}
This equation can easily be solved by the Fourier transformation, giving 
\begin{gather}
\sigma_i(\lambda)=\int_{-\infty}^\infty\frac{d\omega}{2\pi}e^{-i\omega\lambda}\frac{\hat{a}_{1}(\omega,g)}{1+e^{-|\omega|}},
\end{gather}
where
\begin{gather}
\hat{a}_{1}(\omega,g)\equiv \int_{-\infty}^\infty d\lambda\frac{1}{2\pi}\frac{1}{(\lambda+1/g)^2+1/4}e^{i\omega\lambda}.\label{eq_a2Sg}
\end{gather}
The integral \eqref{eq_a2Sg} depends on the Kondo coupling. 
The integrand in Eq.\ \eqref{eq_a2Sg} has two poles located at $\lambda=-1/g\pm i/2$. Therefore, for $0\leq\mathrm{Im}(1/g)<1/2$, we have
\begin{align}
\hat{a}_{1}(\omega,g)=e^{-i\omega/g}(\Theta(\omega)e^{-\omega/2}+\Theta(-\omega)e^{\omega/2}),
\end{align}
and for $\mathrm{Im}(1/g)>1/2$, we have
\begin{align}
\hat{a}_{1}(\omega,g)=\Theta(-\omega)e^{-i\omega/g}(e^{\omega/2}-e^{-\omega/2}),
\end{align}
where $\Theta(\omega)$ is the Heaviside unit-step function. 
Using these results, we obtain the impurity magnetization as $M_i=1/2-\int_{-\infty}^\infty d\lambda \sigma_i(\lambda)=(1-\lim_{\omega\to 0}\hat{a}_1(\omega,g))/2$. We end up with 
$M_i=0$ for $0\leq\mathrm{Im}(1/g)<1/2$ and 
$M_i=1/2$ for $\mathrm{Im}(1/g)>1/2$. Thus, there is a phase transition between the Kondo and the non-Kondo phases at 
\begin{equation}
\mathrm{Im}(1/g)=1/2,
\label{eq_transition}
\end{equation}
accompanied by the jump of the impurity magnetization. In the Kondo phase, the Kondo singlet is formed and the impurity spin is screened. In the non-Kondo phase, the Kondo screening does not occur and the impurity spin remains active.

The transition \eqref{eq_transition} is shown by the red curve in Fig.\ \ref{fig_RGflow}. Remarkably, the exact result shows a good agreement with the RG result in the weak-coupling case $|\rho_0J|\lesssim 0.3$. We can show that the two results exactly coincide in the weak-coupling limit \cite{supple}. The deviation in the strong-coupling case is due to the fact that the Bethe ansatz method requires the linearization of the band dispersion and thus cannot be applied to the strong-coupling case $|\rho_0J|\gtrsim 0.5$ as inferred from the expression of $g$.


\textit{Discussion and conclusion.}--\ 
The inelastic collisions in the alkaline-earth atomic gases are usually considered to be detrimental to observing quantum many-body physics \cite{Riegger18, Hofer, Pagano15}. Nevertheless, here we have shown that the inelastic collisions open a new avenue to non-Hermitian many-body physics. 
Using the previously measured loss rates due to the interorbital inelastic collisions for $^{173}$Yb \cite{Scazza}, we obtain a rough estimate of the imaginary part of the interaction strength as $\rho_0 J_i \sim 10^{-3}$ (here we assume that the hopping rate is of the order of $100$ Hz). This indicates that the atomic gas of $^{173}$Yb is likely to be in the Kondo phase; importantly, we note that the inelastic collision rate can be controlled by external confinement \cite{Riegger18}, an orbital Feshbach resonance \cite{Hofer, Pagano15, ZhangZhaiPeng15}, or photoassociation \cite{Tomita17}. These experimental techniques for controlling the dissipation in atomic gases will enable  detection of the non-Hermitian quantum phase transition. 
We also note that $^{171}$Yb atoms are yet another promising candidate for the non-Hermitian Kondo effect, since an antiferromagnetic spin-exchange interaction has recently been observed \cite{Ono18}, while measurements of the loss rate have been performed only at high temperatures \cite{Ludlow11}. 
The presence of the Kondo state in the atomic gas can be diagnosed by measuring the impurity magnetization and dynamical spin susceptibility \cite{KanaszNagy18}. In addition, the quantum gas microscopy \cite{Ott16} can be used for observing space-resolved spin correlations around the Kondo impurity as well as time-dependent dynamics. 

An important open question is to elucidate an experimental signature of the emergent energy scale $T_{\mathrm{Kdiss}}$ which characterizes the reversion of RG flows. Although there is no clear notion of temperature in the out-of-equilibrium dissipative dynamics, the spatial or temporal evolution of the spin correlations can potentially reflect the characteristics of the RG flow, as in recent numerical results for a Hermitian system \cite{Ashida18_1, Ashida18_2}.

The nature of the non-Hermitian quantum phase transition is also an important issue. The divergent imaginary Kondo interaction in the RG implies that the phase transition is of genuine non-Hermitian nature. Moreover, the Bethe-ansatz method in the thermodynamic limit does not work at the critical point, since the trajectory of the spin rapidity crosses the pole of the integrand in Eq.\ \eqref{eq_a2Sg}. This suggests that the critical point may correspond to an exceptional point \cite{Heiss12}, where the Hamiltonian cannot be diagonalized. This problem merits further study.

The reversion of RG flows discovered in this Letter is not limited to the Kondo effect but can widely emerge when a system has a marginally relevant interaction. 
We thus expect that our finding not only serves as a non-Hermitian generalization of the Kondo physics, but also captures a universal aspect of many-body physics in non-Hermitian quantum systems. The universality of non-Hermitian systems merits future investigation.

\begin{acknowledgments}
We are grateful to Yuto Ashida, Yusuke Horinouchi, Hosho Katsura, Koki Ono, Takafumi Tomita, and Yoshiro Takahashi for helpful discussions. 
This work was supported by KAKENHI (Grants No.\ JP16K05501, No.\ JP18H01140, and No.\ JP18H01145) and a Grant-in-Aid for Scientific Research on Innovative Areas (KAKENHI Grant No.\ JP15H05855) from the Japan Society for the Promotion of Science. 
M.N.\ was supported by RIKEN Special Postdoctoral Researcher Program.
\end{acknowledgments}

\bibliography{inelaKondo_ref.bib}

\begin{thebibliography}{68}
\bibitem [{foo()}]{footnote_perturbativeRG}%
  \BibitemOpen
  \href@noop {} {}\bibinfo {note} {Since the RG equation for the complex
  coupling can be decomposed into RG equations for the real and imaginary
  parts, the perturbative RG is valid if both $|\rho_0J_r|$ and $|\rho_0 J_i|$
  are small.}\BibitemShut {Stop}%
\end{thebibliography}


\clearpage

\renewcommand{\thesection}{S\arabic{section}}
\renewcommand{\theequation}{S\arabic{equation}}
\setcounter{equation}{0}
\renewcommand{\thefigure}{S\arabic{figure}}
\setcounter{figure}{0}

\onecolumngrid
\appendix
\begin{center}
\large{Supplemental Material for}\\
\textbf{``Non-Hermitian Kondo Effect in Ultracold Alkaline-Earth Atoms"}
\end{center}


\section{Detailed derivation of the non-Hermitian Kondo Hamiltonian}

The $s$-wave scattering between the $^1S_0$ state and the $^3P_0$ state is decomposed into two channels \cite{Gorshkov, Scazza} according to whether the two-particle wavefunction is orbital-symmetric spin-singlet
\begin{equation}
\frac{1}{2}(\ket{ge}+\ket{eg})(\ket{\uparrow\downarrow}-\ket{\downarrow\uparrow})\notag
\end{equation}
or orbital-antisymmetric spin-triplet
\begin{gather}
\frac{1}{\sqrt{2}}(\ket{ge}-\ket{eg})\ket{s}\ \mathrm{with}\ 
\ket{s}=\ket{\uparrow\uparrow}, \frac{1}{\sqrt{2}}(\ket{\uparrow\downarrow}+\ket{\downarrow\uparrow}),\ket{\downarrow\downarrow},\notag
\end{gather}
since the total wavefunction of two fermionic atoms should be antisymmetric with respect to exchange of particles. Here, $\ket{g}$ and $\ket{e}$ denote the $^1S_0$ state and the $^3P_0$ state, respectively. 
We denote the annihilation operator of the $^1S_0$ ($^3P_0$) state with spin $\sigma$ by $c_\sigma$ ($f_\sigma$). 
In the second-quantized form, the wavefunctions are expressed as
\begin{align}
c_\uparrow^\dag f_\downarrow^\dag\ket{0} &\leftrightarrow \frac{1}{\sqrt{2}}(\ket{ge}\ket{\uparrow\downarrow}-\ket{eg}\ket{\downarrow\uparrow}),\notag\\
c_\downarrow^\dag f_\uparrow^\dag\ket{0} &\leftrightarrow \frac{1}{\sqrt{2}}(\ket{ge}\ket{\downarrow\uparrow}-\ket{eg}\ket{\uparrow\downarrow}),\notag\\
c_\uparrow^\dag f_\uparrow^\dag\ket{0} &\leftrightarrow \frac{1}{\sqrt{2}}(\ket{ge}-\ket{eg})\ket{\uparrow\uparrow},\notag\\
c_\downarrow^\dag f_\downarrow^\dag\ket{0} &\leftrightarrow \frac{1}{\sqrt{2}}(\ket{ge}-\ket{eg})\ket{\downarrow\downarrow},\notag
\end{align}
and hence
\begin{align}
\frac{1}{\sqrt{2}}(c_\uparrow^\dag f_\downarrow^\dag+c_\downarrow^\dag f_\uparrow^\dag)\ket{0} &\leftrightarrow \frac{1}{2}(\ket{ge}-\ket{eg})(\ket{\uparrow\downarrow}+\ket{\downarrow\uparrow}),\notag\\
\frac{1}{\sqrt{2}}(c_\uparrow^\dag f_\downarrow^\dag-c_\downarrow^\dag f_\uparrow^\dag)\ket{0} &\leftrightarrow \frac{1}{2}(\ket{ge}+\ket{eg})(\ket{\uparrow\downarrow}-\ket{\downarrow\uparrow}).\notag
\end{align}

Assuming that the impurity is located at the origin, the Lindblad operators for two-body losses by the inelastic scattering are given by
\begin{align}
L_\pm &=\sqrt{2\gamma_{eg}^\mp}\frac{1}{\sqrt{2}}(f_\downarrow c_\uparrow(0)\pm f_\uparrow c_\downarrow(0)),\label{eq_Lpm}\\
L_{\uparrow\uparrow}&=\sqrt{2\gamma_{eg}^-}f_\uparrow c_\uparrow(0),\\
L_{\downarrow\downarrow}&=\sqrt{2\gamma_{eg}^-}f_\downarrow c_\downarrow(0)\label{eq_Ldown},
\end{align}
where $c_\sigma(0)$ annihilates an atom in the $^1S_0$ state at the impurity site. The coefficients $\gamma_{eg}^+, \gamma_{eg}^- >0$ are determined by the loss rates due to the inelastic collisions in the two scattering channels. 
Substituting Eqs.\ \eqref{eq_Lpm}-\eqref{eq_Ldown} into the non-Hermitian Hamiltonian, we obtain
\begin{align}
H_{\mathrm{eff}}&=H-\frac{i}{2}\sum_{\alpha=+,-,\uparrow\uparrow,\downarrow\downarrow}L_\alpha^\dag L_\alpha\notag\\
&=H-\frac{i}{2}(\gamma_{eg}^++\gamma_{eg}^-)\sum_{\sigma,\sigma'} c_\sigma^\dag(0) c_\sigma(0) f_{\sigma'}^\dag f_{\sigma'}-\frac{i}{2}(\gamma_{eg}^+-\gamma_{eg}^-)\sum_{\sigma,\sigma'}c_{\sigma}^\dag(0) f_{\sigma'}^\dag c_{\sigma'}(0)f_{\sigma}\notag\\
&=H-\frac{i}{4}(\gamma_{eg}^++3\gamma_{eg}^-)\sum_{\sigma,\sigma'} c_\sigma^\dag(0) c_\sigma(0) f_{\sigma'}^\dag f_{\sigma'}+\frac{i}{2}(\gamma_{eg}^+-\gamma_{eg}^-)\bm{S}_c(0)\cdot\bm{S}_{\mathrm{imp}},
\label{eq_int}
\end{align}
which is the non-Hermitian Kondo Hamiltonian. Here, we define the spin operators as $\bm{S}_c(0)\equiv\frac{1}{2}\sum_{\sigma,\sigma'}c_{\sigma}^\dag(0)\bm{\sigma}_{\sigma\sigma'}c_{\sigma'}(0)$ and $\bm{S}_{\mathrm{imp}}\equiv\frac{1}{2}\sum_{\sigma,\sigma'}f_{\sigma}^\dag\bm{\sigma}_{\sigma\sigma'}f_{\sigma'}$. 
The imaginary interactions $v_i, J_i$ in the non-Hermitian Kondo Hamiltonian in the main text are given by $v_i=-\frac{1}{4}(\gamma_{eg}^++3\gamma_{eg}^-)$ and $J_i=-\frac{1}{2}(\gamma_{eg}^+-\gamma_{eg}^-)$.


\section{Solution of the renormalization-group equation}

In this Appendix, by solving the RG equation (Eq.\ (4) in the main text), we derive the critical line and the energy scale $T_{\mathrm{Kdiss}}$ which characterizes the energy scale at which the reversion of RG flows starts to occur. 
Introducing a dimensionless coupling constant $j=j_r+ij_i\equiv \rho_0J$ ($j_r,j_i\in\mathbb{R}$), we rewrite Eq.\ (4) as
\begin{equation}
\frac{dj}{d\ln D}=j^2+\frac{1}{2}j^3.
\end{equation}
We can integrate this equation from $D_0$ to $D$ as 
\begin{align}
\ln\frac{D}{D_0}&=\int_{j(D_0)}^{j(D)}\frac{dj}{j^2+j^3/2}\notag\\
&=\int_{j(D_0)}^{j(D)}dj\Bigl[\frac{1}{j^2}-\frac{1}{2j}+\frac{1}{4}\frac{1}{1+j/2}\Bigr]\notag\\
&=-\frac{1}{j(D)}-\frac{1}{2}\ln j(D)+\frac{1}{2}\ln(1+\frac{1}{2}j(D))+\frac{1}{j(D_0)}+\frac{1}{2}\ln j(D_0)-\frac{1}{2}\ln(1+\frac{1}{2}j(D_0)).
\label{eq_RGsol_exact}
\end{align}
On the critical line, the coupling constant flows towards $j_r(D)\to-2/3$ and $j_i(D)\to\infty$. 
By comparing the imaginary parts of both sides of Eq.\ \eqref{eq_RGsol_exact} in this limit, 
we obtain a condition for the critical point as
\begin{equation}
\frac{\pi}{2}-\frac{j_i(D_0)}{|j(D_0)|^2}+\frac{1}{2}\arctan\frac{j_i(D_0)}{j_r(D_0)}-\frac{1}{2}\arctan\frac{\frac{1}{2}j_i(D_0)}{1+\frac{1}{2}j_r(D_0)}=0.
\label{eq_RGcrit}
\end{equation}
for $j_i(D_0)>0$. 
In Fig.\ 1, we show the critical line determined by Eq.\ \eqref{eq_RGcrit} by using the blue curve.

Next, we derive the energy scale $T_{\mathrm{Kdiss}}$ defined by $j_r(T_{\mathrm{Kdiss}})=0$ (and $j_i(T_{\mathrm{Kdiss}})> 0$) for the case of $j_r(D_0)<0$. To derive the expression of $T_{\mathrm{Kdiss}}$, we set $j_r(D)\to -0$ in Eq.\ \eqref{eq_RGsol_exact} and take its imaginary part. Then, we obtain
\begin{align}
0=\frac{\pi}{4}+\frac{1}{j_i(D)}+\frac{1}{2}\arctan(\frac{1}{2}j_i(D))+\frac{-j_i(D_0)}{|j(D_0)|^2}+\frac{1}{2}\arctan\frac{j_i(D_0)}{j_r(D_0)}-\frac{1}{2}\arctan\frac{\frac{1}{2}j_i(D_0)}{1+\frac{1}{2}j_r(D_0)},
\label{eq_gi}
\end{align}
which determines $j_i(D)$. 
Note that $j_i(D)\to\infty$ when the condition \eqref{eq_RGcrit} is satisfied. 
Taking the real part of Eq.\ \eqref{eq_RGsol_exact} and set $D=T_{\mathrm{Kdiss}}$, we get
\begin{align}
&\ln\frac{T_{\mathrm{Kdiss}}}{D_0}=-\frac{1}{2}\ln \tilde{j}_i+\frac{1}{2}\ln\sqrt{1+\frac{1}{4}\tilde{j}_i^2}+\frac{j_r(D_0)}{|j(D_0)|^2}+\frac{1}{2}\ln|j(D_0)|-\frac{1}{2}\ln |1+\frac{1}{2}j(D_0)|,
\end{align}
where $\tilde{j}_i\equiv j_i(T_{\mathrm{Kdiss}})$ is determined by Eq.\ \eqref{eq_gi}. 
Therefore, we arrive at
\begin{align}
T_{\mathrm{Kdiss}}/D_0=\tilde{j}_i^{-1/2}(1+\frac{1}{4}\tilde{j}_i^2)^{1/4}|j(D_0)|^{1/2}|1+\frac{1}{2}j(D_0)|^{-1/2}\exp\left[\frac{j_r(D_0)}{|j(D_0)|^2}\right],
\end{align}
which is equivalent to Eq.\ (5) in the main text.


\section{Critical line in the weak-coupling limit}

In this Appendix, we derive the critical line in the weak-coupling ($|\rho_0J|\ll 1$) case. Using this, we compare the RG result with the exact Bethe-ansatz solution. 
First, we consider the RG solution for the critical line. Since we can read off that $j_i(D_0)/j_r(D_0)\to 0$ with $j_i(D_0)\to 0$ on the critical line, we can expand Eq.\ \eqref{eq_RGcrit} using $|j_i(D_0)/j_r(D_0)|\ll 1$ and $|j(D_0)|\ll 1$. Taking the lowest-order contribution, we obtain
\begin{align}
&j_r(D_0)^2+\Bigl(j_i(D_0)-\frac{1}{\pi}\Bigr)^2=\frac{1}{\pi^2}.\label{eq_circle}
\end{align}
Therefore, the critical line approaches a circle in the weak-coupling limit.

On the other hand, in the Bethe-ansatz solution, the transition point from the Kondo phase to the non-Kondo phase is determined by the condition
\begin{gather}
\mathrm{Im}(1/g)=1/2,
\end{gather}
where $g=-\tan(\pi\rho_0 J)$. 
This condition is rewritten as
\begin{align}
&\mathrm{Im}(1/g)=\frac{\sinh(2\pi j_i)}{\cosh(2\pi j_i)-\cos(2\pi j_r)}=1/2\notag\\
\Leftrightarrow\ &\sinh(2\pi j_i)-\frac{1}{2}\cosh(2\pi j_i)=-\frac{1}{2}\cos(2\pi j_r).
\label{eq_Bethecrit}
\end{align}
When $|j_r|, |j_i|\ll 1$, by expanding the both sides of the above equation, we obtain
\begin{gather}
2\pi j_i-\frac{1}{2}-\frac{(2\pi j_i)^2}{4}=-\frac{1}{2}+\frac{(2\pi j_r)^2}{4}\notag\\
\Leftrightarrow\ j_r^2+\Bigl(j_i-\frac{1}{\pi}\Bigr)^2=\frac{1}{\pi^2},
\end{gather}
which agrees with the RG result in Eq.\ \eqref{eq_circle}.

\end{document}